\documentclass[a4paper,11pt]{article}
\usepackage{pos}

\title{Gluon tomography through diffractive processes in a saturation framework}

\author[a]{Renaud Boussarie}
\author*[b]{Michael Fucilla}
\author[c,d]{Andrey V. Grabovsky}
\author[b]{Emilie Li}
\author[e]{Lech Szymanowski}
\author[b]{Samuel Wallon}

\affiliation[a]{CPHT, CNRS, Ecole Polytechnique, Institut Polytechnique de Paris, 91128 Palaiseau, France}

\affiliation[b]{Université Paris-Saclay, CNRS/IN2P3, IJCLab, 91405, Orsay, France}

\affiliation[c]{Budker Institute of Nuclear Physics, 11, Lavrenteva avenue, 630090, Novosibirsk, Russia}

\affiliation[d]{Novosibirsk State University, 630090, 2, Pirogova street, Novosibirsk, Russia}

\affiliation[e]{National Centre for Nuclear Research (NCBJ), Pasteura 7, 02-093 Warsaw,  Poland}

\emailAdd{Renaud.Boussarie@polytechnique.edu}
\emailAdd{Michael.Fucilla@unical.it}
\emailAdd{A.V.Grabovsky@inp.nsk.su}
\emailAdd{Emilie.Li@ijclab.in2p3.fr}
\emailAdd{Lech.Szymanowski@ncbj.gov.pl}
\emailAdd{Samuel.Wallon@ijclab.in2p3.fr}

\abstract{We discuss a series of results aimed at bringing saturation physics and gluon tomography into an era of precision. In particular, the NLO treatment of diffractive: 1)  exclusive dijet, 2)  exclusive longitudinally polarized light vector meson and 3) semi-inclusive single or double hadron photo- or electroproduction with large $p_T$, on a nucleon or a nuclei. Finally, we discuss the more complicated 4) exclusive transversely polarized light vector meson production, which starts at the next-to-leading power and therefore requires a beyond leading twist treatment. This new class of processes provides an access to precision physics of gluon saturation dynamics, with very promising future phenomenological studies at the EIC, or, at the LHC in $p A$ and $A A$ scattering, using Ultra Peripheral Collisions (UPC).}

\FullConference{31st International Workshop on Deep Inelastic Scattering (DIS2024)\\
 8–12 April 2024\\
Grenoble, France\\}

\begin{document}
\maketitle

\section{Introduction}
The investigation of strong interactions at very high-energy has been the object of intense studies for many decades. Particularly interesting is the so-called Regge-Gribov (or semi-hard) limit of QCD, characterized by the scale hierarchy $s \gg Q^2 \gg \Lambda_{ \rm QCD}^2$, where $\sqrt{s}$ is the center-of-mass energy, $Q$ is  a hard scale characterizing the process and $\Lambda_{ \rm QCD}$ is the QCD mass scale. This limit is the stage where some of the most intriguing phenomena of Quantum Chromodynamics (QCD) manifest themselves, such as the saturation of the gluon density inside the proton with the subsequent formation of a state of hadronic matter characterized by a high density of particles and disordered field distribution, that is known by the name of color glass condensate (CGC). The experimental investigation of CGC is an intriguing opportunity at modern accelerators, such as the Large Hadron Collider (LHC) and the Relativistic Heavy Ion Collider (RHIC), but, more importantly, constitutes one of the pillars of the forthcoming Electron-Ion Collider (EIC) physics program. \\

\noindent \textbf{Semi-classical small-$x$ EFT} \\

\noindent In the saturation regime the target becomes a highly dense many-body system, whose evolution is driven by non-linear dynamics. The best option for dealing with such a complicated system is to develop an effective field theory. The key observation that the relevant degrees of freedom at small-$x$ are classical~\cite{McLerran:1993ni}, led to the development of such effective field theories~\cite{McLerran:1993ka,McLerran:1994vd,Balitsky:2001re}. Introducing a light-cone basis using vectors $n_1$ (projectile direction) and $n_2$ (target direcion), which specify the $+/-$ directions and such that any four-vector can be written
\begin{equation}
    k = k^+ n_1^{\mu} + k^{-} n_2^{\mu} + k_{\perp} \; , 
\end{equation}
the gluonic field $A$ is separated into external classical background fields $b$ and internal quantum fields $\mathcal{A}$. The separation is made depending on whether their $+$-momentum is above or below the arbitrary rapidity cut-off $e^\eta p_\gamma^+$, with $\eta < 0$. The external field, after being highly boosted from the target rest frame to the probe frame, takes the form 
\begin{equation}
    b^\mu (x) = b^-(x_\perp) \delta (x^+) n_2^\mu \,.
\end{equation}
The resummation of all order interactions with those fields leads to a high-energy Wilson line, that represents a shockwave and is located exactly at $z^- =0$:
\begin{equation}
    U_{\vec{z}} = \mathcal{P} \exp \left(i g \int d z^+ b^-(z)\right)\,,
\end{equation}
where $\mathcal{P}$ is the usual path ordering operator for the $+$ direction. Matrix elements constructed from the high-energy Wilson lines describe the dense target. Their renormalization group equations with respect to the rapidity cut-off $\eta$ allow for the resummation of small-$x$ logarithms~\cite{Balitsky:1995ub,Jalilian-Marian:1997qno,Jalilian-Marian:1997jhx,Jalilian-Marian:1997ubg,Jalilian-Marian:1998tzv,Weigert:2000gi,Iancu:2000hn,Iancu:2001ad}. Relying on the small-$x$ factorization, the scattering amplitude can be written as the convolution of the projectile impact factor with the non-perturbative matrix element of operators from the Wilson line operators acting on the target states. 

\section{Diffractive processes at the NLO}
\label{NLO}

\noindent The semi-classical small-$x$ effective theory, when applied to diffractive processes, represents an excellent tool to investigate the five dimensional Wigner distribution~\cite{Marquet:2009ca,Hatta:2022lzj}. Excellent probes of this latter are: (\emph{i}.) the diffractive dijet production, $\gamma^{*} P \rightarrow \rm jet_1 + jet_2 + X + P'$, (\emph{ii}.) the semi-inclusive diffractive deep inelastic scattering (SIDDIS), $\gamma^{*} P \rightarrow h + X + P'$, (\emph{iii}.) the diffractive dihadron production, $ \gamma^{*} P \rightarrow h_1 + h_2 + X + P'$ and (\emph{iv}.) the deeply virtual meson production (DVMP), $\gamma^{*} (\lambda_{\gamma}) P \rightarrow M (\lambda_{\gamma}) P$. The full NLO treatment of (\emph{i}.), (\emph{ii}.) and (\emph{iii}.) have been obtained in~\cite{Boussarie:2014lxa,Boussarie:2016ogo,Fucilla:2022wcg,Fucilla:2023mkl}. For (\emph{iv}.), NLO calculations exist only in the case of longitudinally polarized vector meson~\cite{Boussarie:2016bkq,Mantysaari:2022bsp,Ivanov:2004pp}. \\

\begin{figure}
\centering
 \includegraphics[scale=0.28]{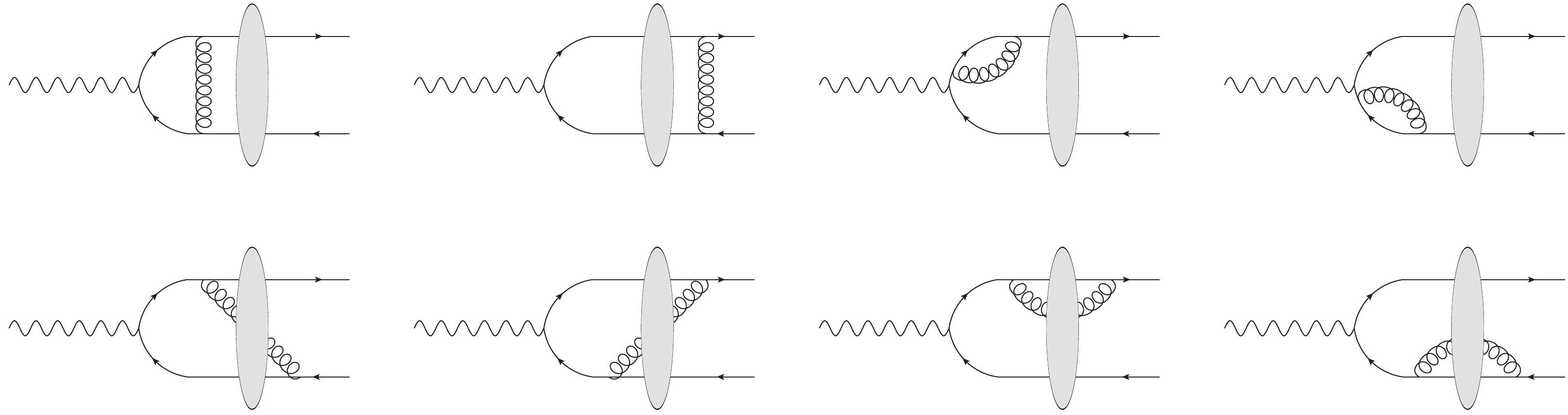}
 \caption{One loop diagrams for $\gamma^{*} \rightarrow q \bar{q}$.}
    \label{fig:OneLoopDipole}
\end{figure}

\begin{figure}
\centering
 \includegraphics[scale=0.28]{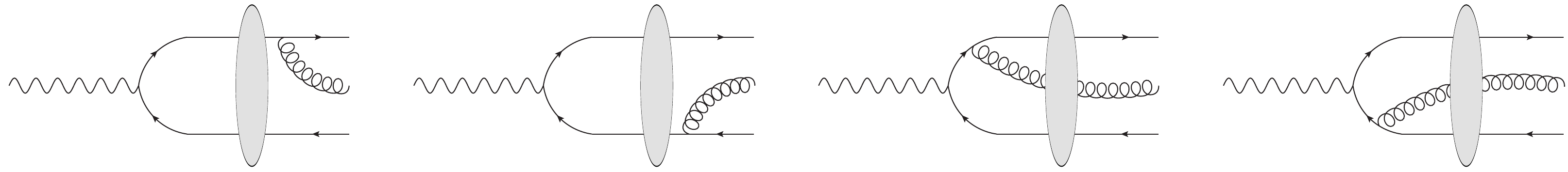}
 \caption{Born diagrams for $\gamma^{*} \rightarrow q \bar{q} g$.}
    \label{fig:BornDipolePlusg}
\end{figure}

\noindent \textbf{Next-to-leading order corrections to the $\gamma^{*} \rightarrow q \bar{q} $ impact factor.} \vspace{0.2 cm} \\ 
The calculation of the cross sections of the aforementioned processes, within next-to-leading order accuracy, requires:
\begin{itemize}
    \item[\textbullet] The one-loop $\gamma^{*} \rightarrow q \bar{q} $ impact factor, which enters the virtual corrections. Diagrams contributing to this latter are shown in fig.~\ref{fig:OneLoopDipole}. 
    \item[\textbullet] The $\gamma^{*} \rightarrow q \bar{q} g$ impact factor at the Born level, which enters the real corrections. Diagrams contributing to this latter are shown in fig.~\ref{fig:BornDipolePlusg}.
\end{itemize}
In the diffractive case, i.e. when a color singlet is exchanged in the $t$-channel at the amplitude level, these results where obtain for the first time in ref.~\cite{Boussarie:2014lxa,Boussarie:2016ogo}. \\

\noindent \textbf{IR-sector and rapidity divergences} \vspace{0.2 cm}  \\
At the NLO, we encounter several contributions that are separately singular and, for this reason, the divergences must first be regularized to 
demonstrate their explicit cancellation. Divergences are easily classified when the Sudakov decomposition for the momenta  is used:
\begin{equation}
\label{p-sudakov}
p_p^\mu = x_p p_{\gamma}^+ n_1^\mu + \frac{\vec{p}_p^{\,2}}{2 x_p p_{\gamma}^+} n_2^\mu + p_{p,\perp}^\mu\,,
\end{equation} 
where $p=q,\bar{q},g$ for quark, anti-quark and gluon, respectively. Concerning the IR-sector, we encounter 
\begin{itemize}
    \item \textbf{Rapidity divergences}: $x_g \rightarrow 0$ while $p_{g, \perp}$ is fixed but strongly suppressed with respect to $p_{\gamma}^{+} \sim \sqrt{s}$.
    \item \textbf{Collinear divergences}: $p_{g, \perp} \rightarrow (x_g/x_q) p_{q, \perp}$ (collinear to the quark line) or $p_{g, \perp} \rightarrow (x_g/x_{\bar{q}}) p_{\bar{q}, \perp}$ (collinear to the anti-quark line) while $x_g$ is generic.
    \item \textbf{Soft divergences}: All components linearly vanishing (both $x_g$ and $p_{g, \perp}$ go linearly to zero). Parameterizing the transverse momenta of the gluon as $p_{g, \perp} = x_g u_{\perp}$, with $|u_{\perp}|$ fixed in the limit $x_g$ goes to zero, we can then define the soft limit as $x_g$ goes to zero with $u_{\perp}$ generic. 
    \item \textbf{Soft and collinear divergences}: Soft as defined above, but with $u_{\perp} \rightarrow (1/x_q) p_{q, \perp}$ (soft and collinear to the quark line) or $u_{\perp} \rightarrow (1/x_{\bar{q}}) p_{\bar{q}, \perp}$ (soft and collinear to the anti-quark line).
\end{itemize}
To regularize the various singularities, we introduce a lower cut-off in the variable $x_g$, i.e. we set $|x_g| < \alpha$, and use dimensional regularization for the transverse components ($d = D-2= 2+2 \epsilon$). Because of loop corrections, in the virtual contributions, we also face 
\begin{itemize}
    \item[\textbullet] \textbf{Ultraviolet singularities}: $p_{g}^2 \rightarrow \infty$. When the Sudakov decomposition is used and the component along the four-vector $n_2$ is integrated through the Cauchy theorem, UV-divergences can only manifest themselves in the limit $p_{g \perp}^2 \rightarrow \infty$.  \\
\end{itemize}

The cancellation of UV and rapidity divergences can be shown at the amplitude level and it is therefore common to all processes. In axial gauges, UV renormalization is commonly tedious. In the present case, many complications can be avoided by setting $\epsilon \equiv \epsilon_{\rm UV} = \epsilon_{ \rm IR}$. In fact, the only UV singularities of our impact factors are related with the dressing of final quark states\footnote{Running coupling effects are excluded from the projectile and completely factorized in the target part of the cross-section}. These latter have the form of tadpole integrals vanishing when $\epsilon \equiv \epsilon_{ \rm UV} = \epsilon_{\rm IR}$. Virtual diagrams of fig.~\ref{fig:OneLoopDipole}, more specifically the ones in which the gluon crosses the Shockwave, contain rapidity divergences. The rapidity
divergent-terms have to be absorbed into the renormalized Wilson operators with the help
of the B-JIMWLK equation. We thus have to use the B-JIMWLK evolution for these
operators, in the leading amplitude, from a aribitrary cutoff $\alpha$ to the rapidity divide $e^{\eta}$, to produce  counterterms that cancel the rapidity divergences. \\ 

The cancellation of infrared divergences can be observed only at the cross section level and it is process-dependent. Soft divergences can be isolated through a subtraction procedure and cancel in the real plus virtual combination. In the dijet case, the collinear singularities cancel when a proper NLO jet algorithim is used. In the semi-inclusive hadrons production and in the exclusive light vector meson production, the renormalization of the NLO fragmentation functions and distibution amplitudes, respectively, removes the remaining collinear singularities.   

\section{Higher-twist corrections to the DVMP at the LO}
\label{Higher-twist}

\noindent Besides NLO contributions, there are other types of corrections which are important for the phenomenology of diffractive processes: the higher-twist and the subeikonal corrections (see for instance~\cite{Altinoluk:2014oxa}). These allow, on the one hand, to increase the precision of every possible observable and, on the other, to consider physical effects which are invisible under too restrictive approximations. This is for instance the case of target spin effects~\cite{Kovchegov:2019rrz}, which vanish in the eikonal approximation and require a subeikonal treatment. \\

Another emblematic example is the DVMP in the case of a transversely polarized light vector meson. This process vanishes at the twist 2, to all order in perturbation theory, and therefore calls for a beyond leading twist treatment. In the saturation regime, this process has been considered in ref.~\cite{Boussarie:2024bdo} through a method combining the higher-twist formalism of exclusive processes in the $s$ channel with the semi-classical effective description of small-$x$ physics in the $t$ channel~\cite{Boussarie:2024pax}. An intriguing future perspective is to combine the technologies described in this section with that of section \ref{NLO}, to obtain the complete NLO description of the DVMP at the twist 3. Although the frameworks described above are conceptually simple to combine, a complete NLO calculation requires one-loop corrections to the $\gamma^{*} \rightarrow q \bar{q} g$ vertex in fig.~\ref{fig:BornDipolePlusg}, which is extremely challenging and yet unknown. 


\begin{thebibliography}{99}

\bibitem{McLerran:1993ni}
L.~D.~McLerran and R.~Venugopalan,
Phys. Rev. D \textbf{49} (1994), 2233-2241
[arXiv:9309289 [hep-ph]].

\bibitem{McLerran:1993ka}
L.~D.~McLerran and R.~Venugopalan,
Phys. Rev. D \textbf{49} (1994), 3352-3355
[arXiv:9311205 [hep-ph]].

\bibitem{McLerran:1994vd}
L.~D.~McLerran and R.~Venugopalan,
Phys. Rev. D \textbf{50} (1994), 2225-2233
[arXiv:9402335 [hep-ph]].

\bibitem{Balitsky:2001re}
I.~Balitsky,
Phys. Lett. B \textbf{518} (2001), 235-242
[arXiv:0105334 [hep-ph]].

\bibitem{Balitsky:1995ub}
I.~Balitsky,
Nucl. Phys. B \textbf{463} (1996), 99-160
[arXiv:9509348 [hep-ph]].

\bibitem{Jalilian-Marian:1997qno}
J.~Jalilian-Marian, A.~Kovner, A.~Leonidov and H.~Weigert,
Nucl. Phys. B \textbf{504} (1997), 415-431
[arXiv:9701284 [hep-ph]].

\bibitem{Jalilian-Marian:1997jhx}
J.~Jalilian-Marian, A.~Kovner, A.~Leonidov and H.~Weigert,
Phys. Rev. D \textbf{59} (1998), 014014
[arXiv:9706377 [hep-ph]].

\bibitem{Jalilian-Marian:1997ubg}
J.~Jalilian-Marian, A.~Kovner and H.~Weigert,
Phys. Rev. D \textbf{59} (1998), 014015
[arXiv:9709432 [hep-ph]].

\bibitem{Jalilian-Marian:1998tzv}
J.~Jalilian-Marian, A.~Kovner, A.~Leonidov and H.~Weigert,
Phys. Rev. D \textbf{59} (1999), 034007
[erratum: Phys. Rev. D \textbf{59} (1999), 099903]
[arXiv:9807462 [hep-ph]].

\bibitem{Weigert:2000gi}
H.~Weigert,
Nucl. Phys. A \textbf{703} (2002), 823-860
[arXiv:0004044 [hep-ph]].

\bibitem{Iancu:2000hn}
E.~Iancu, A.~Leonidov and L.~D.~McLerran,
Nucl. Phys. A \textbf{692} (2001), 583-645
[arXiv:hep-ph/0011241 [hep-ph]].

\bibitem{Iancu:2001ad}
E.~Iancu, A.~Leonidov and L.~D.~McLerran,
Phys. Lett. B \textbf{510} (2001), 133-144
[arXiv:hep-ph/0102009 [hep-ph]].

\bibitem{Marquet:2009ca}
C.~Marquet, B.~W.~Xiao and F.~Yuan,
Phys. Lett. B \textbf{682} (2009), 207-211
[arXiv:0906.1454 [hep-ph]].

\bibitem{Hatta:2022lzj}
Y.~Hatta, B.~W.~Xiao and F.~Yuan,
Phys. Rev. D \textbf{106} (2022) no.9, 094015
[arXiv:2205.08060 [hep-ph]].

\bibitem{Boussarie:2014lxa}
R.~Boussarie, A.~Grabovsky, L.~Szymanowski, and S.~Wallon, 
{\em JHEP} {\bf 1409} (2014) 026, 
[arXiv:1405.7676 [hep-ph]].

\bibitem{Boussarie:2016ogo}
R.~Boussarie, A.~V. Grabovsky, L.~Szymanowski, and S.~Wallon,  
{\em JHEP} {\bf 11} (2016) 149,
[arXiv:1606.00419 [hep-ph].

\bibitem{Fucilla:2022wcg}
M.~Fucilla, A.~V.~Grabovsky, E.~Li, L.~Szymanowski and S.~Wallon,
JHEP \textbf{03} (2023), 159,
[arXiv:2211.05774 [hep-ph]].

\bibitem{Fucilla:2023mkl}
M.~Fucilla, A.~Grabovsky, E.~Li, L.~Szymanowski and S.~Wallon,
JHEP \textbf{02} (2024), 165
[arXiv:2310.11066 [hep-ph]].

\bibitem{Boussarie:2016bkq}
R.~Boussarie, A.~V. Grabovsky, D.~{\relax Yu}. Ivanov, L.~Szymanowski, and S.~Wallon,  {\em Phys. Rev. Lett.} {\bf 119} (2017), no.~7 072002, 
[arXiv:1612.08026].

\bibitem{Mantysaari:2022bsp}
H.~M\"antysaari and J.~Penttala,
Phys. Rev. D \textbf{105} (2022) no.11, 114038
[arXiv:2203.16911 [hep-ph]].

\bibitem{Ivanov:2004pp}
D.~Y.~Ivanov, M.~I.~Kotsky and A.~Papa,
Eur. Phys. J. C \textbf{38} (2004), 195-213
[arXiv:0405297 [hep-ph]].

\bibitem{Altinoluk:2014oxa}
T.~Altinoluk, N.~Armesto, G.~Beuf, M.~Mart\'\i{}nez and C.~A.~Salgado,
JHEP \textbf{07} (2014), 068
[arXiv:1404.2219 [hep-ph]].

\bibitem{Kovchegov:2019rrz}
Y.~V.~Kovchegov,
JHEP \textbf{03} (2019), 174
[arXiv:1901.07453 [hep-ph]].

\bibitem{Boussarie:2024bdo}
R.~Boussarie, M.~Fucilla, L.~Szymanowski and S.~Wallon,
[arXiv:2407.18115 [hep-ph]].

\bibitem{Boussarie:2024pax}
R.~Boussarie, M.~Fucilla, L.~Szymanowski and S.~Wallon,
[arXiv:2407.18203 [hep-ph]].

\end{thebibliography}
\end{document}